\documentclass{amsart}
\usepackage{mathptmx}
\usepackage{amssymb}
\usepackage{amsmath, amsthm, amssymb}
\usepackage{multirow}
\usepackage[english]{babel}
\usepackage{graphicx}
\usepackage{dcolumn}
\usepackage{bm}
\usepackage{hyperref}

\begin{document}

\title[energy saving in Fish school]{Using Robotic Fish to Explore the Hydrodynamic Mechanism of Energy Saving in a Fish School}

\author[L. Li]{Liang Li}
\address{Intelligent Control Laboratory, College of Engineering, Peking University, Beijing 100871, P.~R.~China}
\email{liatli@pku.edu.cn}

\author[L. Jia]{Lichao Jia}
\address{State Key Laboratory for Turbulence and Complex Systems, College of Engineering , Peking University, Beijing 100871, P.~R.~China}
\email{lchjia@pku.edu.cn}

\author[G. Xie]{Guangming Xie}
\address{Intelligent Control Laboratory, College of Engineering, Peking University, Beijing 100871, P.~R.~China}
\email{xiegming@pku.edu.cn}




\begin{abstract}
Fish often travel in highly organized schools.
One of the most quoted functions of these configurations is energy savings.
Here, we verified the hypothesis and explored the mechanism through series of experiments on ``schooling" robotic fish, which can undulate actively with flexible body, resembling real fish.
We find that, when the school swims in the same spatial arrays as the real one, the energy consumption of the follower mainly depends on the phase difference, a phase angle by which the body wave of the follower leads or lags that of the leader, instead of spatial arrays.
Further analysis through flow visualization indicates that the follower saves energy when the phase difference corresponds to the situation that the follower flaps in the same direction of
the flow field induced by the vortex dipole shedding by the leader.
Using biomimetic robots to verify the biological hypothesis in this paper also sheds new light on the connections among the fields of engineering, physics and biology.
\end{abstract}

\maketitle
\section{Introduction}

In the animal kingdom, many species move in highly organized formations \cite{Krause2002Living,Couzin2003Self-organization,Sumpter2010Collective}.
Some well-known systems include the queuing of migrating lobsters \cite{Bill1976Drag}, organized swarming ducklings \cite{Fish1995Kinematics},  flocking birds \cite{Steven2014Upwash}, and
schooling fishes \cite{Partridge1980Three-dimensional,Marras2014Fish}.
Apart from the social benefits \cite{Krause2002Living,Miller2013Both,Berdahl2013Emergent} and physiological factors \cite{Breder1951Studies}, energy saving for individuals is another prevalent
hypothesis of why animals prefer to move in groups.
In the case of fish schools, since Weihs first predicted that fish swimming in school will save energy \cite{Weihs1973Hydromechanics}, this hypothesis has been proved through series of methods, such as theoretical analysis \cite{Weihs1975Some}, computational methods \cite{Dong2007Characteristics,Kanso2010Locomotory,Hemelrijk2014increased,Zhu2014Flow-mediated}, biological experiments
experiments \cite{Svendsen2003Intra,Killen2012Aerobic,Liao2003Fish,Marras2014Fish} and physical models
\cite{Ristroph2008Anomalous,Dewey2014Propulsive,Boschitsch2014Propulsive}.
For theoretical analysis and computational methods, turbulent flow is difficult to estimate.
These methods can only draw an approximate conclusion.
For experiments on real fish, it is difficult to control the fish school and decouple the social factors from physical factors.
Recently, with the development of mechanics, electronics, materials and controls, it is possible to build a man-made "fish school" to unravel the hydrodynamic mechanics of fish school.
Previous physical models were passive flapping filaments that shed different vortices from real fish \cite{Ristroph2008Anomalous,Jia2008Passive}, or, rigid foils that have no body shape or body wave \cite{Dewey2014Propulsive,Boschitsch2014Propulsive}.
Using a high-fidelity fish-like robot to uncover the hydrodynamic mechanism of the fish school has never been explored.

In previous studies, on the energy harvest of fish school and energy expenditure was mainly concerned in relation to lattices \cite{Breder1976Fish}.
Among the formation hypotheses, a prevalent example is the diamond lattice first proposed by Weihs \cite{Weihs1973Hydromechanics}.
Recent studies, based on computational model, predicted that group configurations including diamond, rectangular, phalanx and line are all hydrodynamically advantageous \cite{Hemelrijk2014increased,Zhu2014Flow-mediated}.
Further studies on the real fish school demonstrated energy savings regardless of individuals' spatial positions \cite{Marras2014Fish}.
Whether spatial location is a crucial factor for energy expenditure when fish are swimming in groups is still controversial.
Other factors related to energy consumption remain largely unexplored.

Here, we consider a school of two fish-like robots swimming in a laminar flow.
The high-fidelity of our robotic fish resides in its body shape, bio-inspired locomotion and similar Reverse Karman Vortexes (RKV) shedding by the caudal fin.
The flow first encounters the rigid head and then the flexible body.
Thus, a realistic body shape is critical.
The locomotion should be similar to the real fish to generate a similar flow field.
Moreover, since the energy transfer between fish in school is mostly based on the vortices near the caudal fin, thus a similar wake is also significant.

In this study, we test and compare the power consumed by two robot flapping fish fixed in a flow tunnel at different relative positions and flapping phase differences.
To make fish body undulate in the flow field, the robot should consume power to overcome the resistive force of water.
When robots condense into a group state, water-mediated interactions may lead to modifications of energy cost to keep robots undulating in the same body wave.
Through series of experiments, we find that (a) the trailing fish will benefit energy saving when swimming in some conditions, and (b) when the schooling robotic fish swim in the similar spatial arrays of the real fish, the energy cost of the trailing fish mostly depends on the phase difference.

\begin{figure}[!htb]
  \centering
  \includegraphics[width=0.8\linewidth]{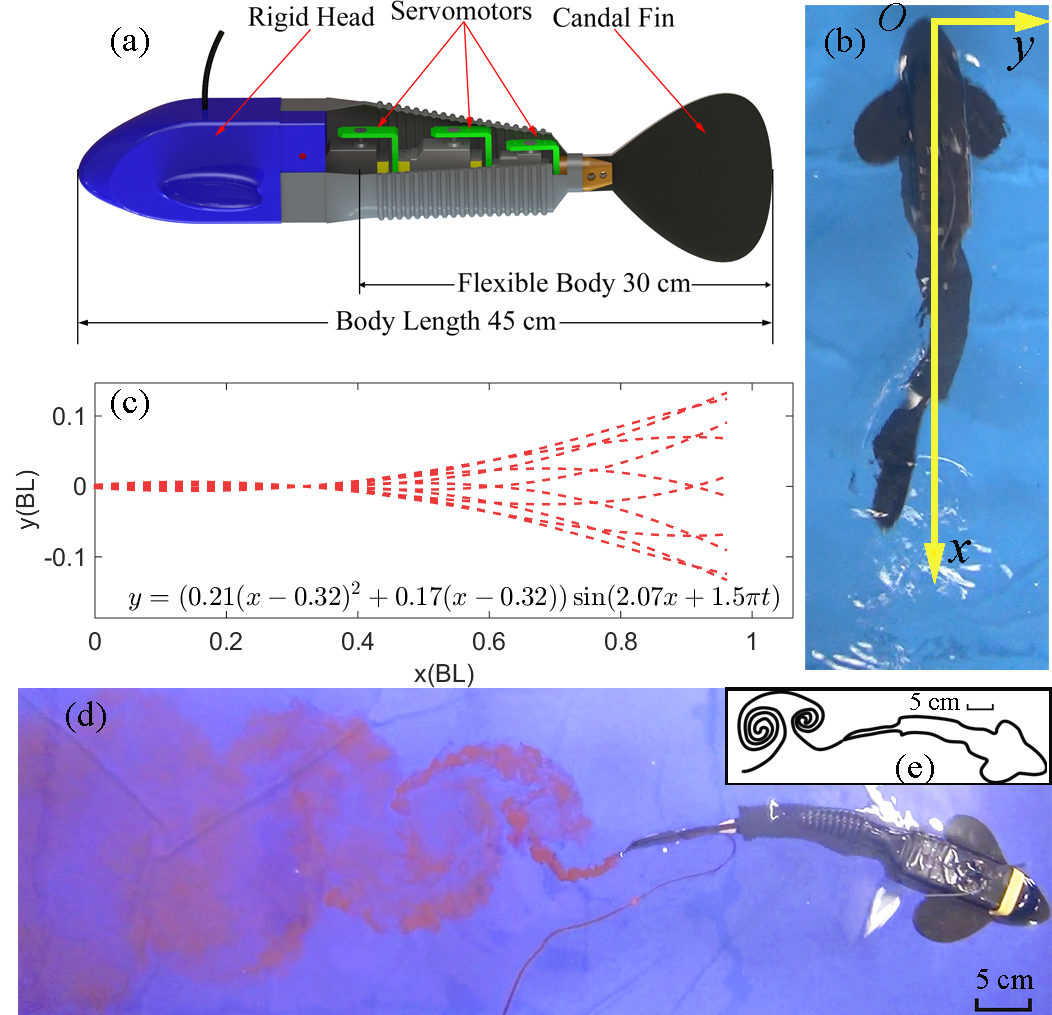}\\
  \caption{ An introduction of the robotic fish, including (a) three-dimensional mechanism consisting of rigid head, flexible body and caudal fin,  (b) a snapshot of
  robotic fish swimming in water in the top view,  (c) a cluster of body undulation derived from real carangiform fish, where $BL$ represents body length, (d) a snapshot of vortices shedding
  by the robotic fish while swimming forward, (e) a typical schematic of the vortices. (online version in color.)}\label{FigBodyAndWaveAndFV}
\end{figure}

\section{Material and methods}
\subsection{The high-fidelity robotic fish}
The robotic fish used in our experiments is designed according to a kind of carangiform fish, which propel themselves by the undulation of the rear body and caudal fin.
In order to improve the fidelity of the flow-mediated interactions between schooling fish, we enable our robotic fish to (a) have a controllable flexible body, (b) swim with biomimetic locomotion and (c) shed a similar flow field after the tail.

The first step is to endow the robotic fish with a fish-like body.
Based on the analyses of biologists, the flexible body has two almost mutually exclusive mechanical properties for swimming: flexible in lateral directions and incompressible in the longitudinal direction \cite{Videler1993FishSwimming}.
In order to meet these requirements, we construct our robotic fish with three servomotors to generate the lateral rhythmic locomotion.
The connections between them are rigid aluminium frames, in longitudinal director.
The caudal fin is made with stiffness decreasing from the peduncle to the tail end, resembling to the real one.

Most carangiform fish swim forward by passing a body wave down their bodies, a mode known as undulatory propulsion.
One of the widely used body wave function is proposed by Lighthill \cite{Lighthill1960Note}.
Here we modified the function as \cite{Liang2014Modeling}.
\begin{equation}
  y = (c_1(x-x_0)+c_2(x-x_0)^2)\sin(kx+2\pi ft)
\end{equation}
where $x$ is the displacement along the boy axis and $y$ is the lateral body displacement (see Figure. \ref{FigBodyAndWaveAndFV} (b)), $c_1$ and $c_2$ are linear and quadratic wave 
amplitude envelope, $k$ denotes the number of whole body wave length, $f$ is the frequency of body wave, $x_0$ is defined as fixed point, which has no oscillation in $y$ axis.
The pattern of body undulation is shown in Figure. \ref{FigBodyAndWaveAndFV}(c).
From the body wave function, we find that each segment of the fish body has its own oscillating amplitude as well as coupled phase differences with its adjacent segments.
Guided by the body wave undulation, we implemented a similar undulation mode in our robotic fish by a bio-inspired controller, Central Pattern Generator (CPG) controller  \cite{CPGreview,SM,Liang2014General}.
The advantages of using CPG controller are that both the phase difference among the joints and between the fish can be well controlled on line.  

Apart from a shape-similar body and bimimetic locomotion, the flow field shedding by the robotic fish is also critical for the interactions between ``schooling" robotic fish.
If we can generate a similar flow field around the robotic fish to the real one, the energy consumption of real fish school can be well estimated by our robotic fish.
We compared the vortex shedding by real fish \cite{Aleyev1977Nekton} and our robotic fish to show the similarity between them \cite{SM}.
From the snapshot of the water flow of robotic fish in Figure. \ref{FigBodyAndWaveAndFV} (d), we can find vortices shed to the left-hand side of the robot rotate clockwise, and vortices shed to the right-hand side rotate counterclockwise  (see Figure. \ref{FigBodyAndWaveAndFV} (e)).
This is quite similar to the Reverse Karman Vortex Street (RKVS), by which fish generates thrusts.

\begin{figure}[!ht]
  \centering
  \includegraphics[width=0.8\linewidth]{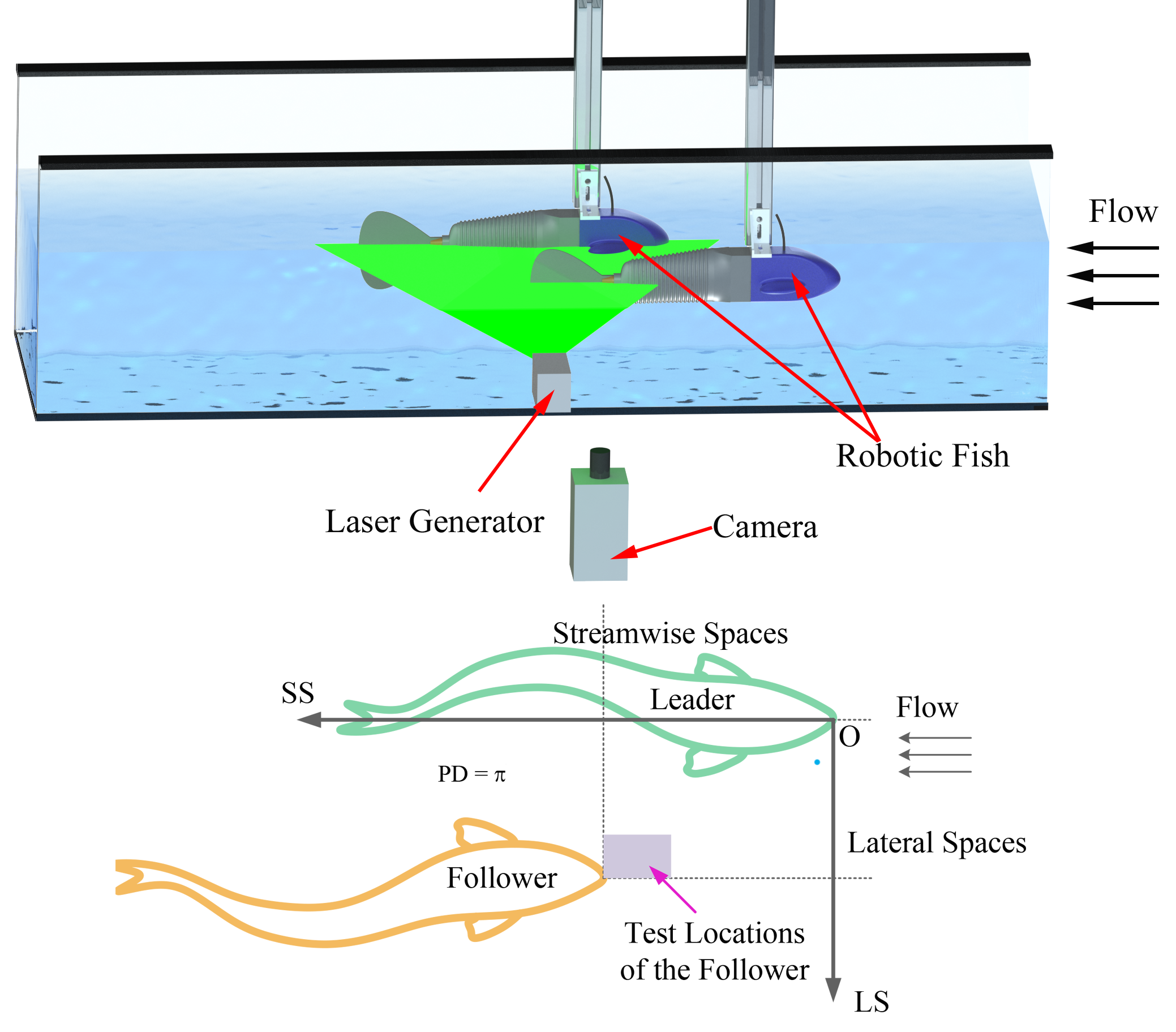}\\
  \caption{An introduction of the test apparatus and the factors. (a) Lateral view of the experimental set-up.  Two robotic fish are fixed in a laminar flow with a free surface.  A camera captures the interactions among the two robotic fish and the vortices. (b) Top view of fish swimming in group with Lateral Space (LS), Streamwise Space (SS) and anti-phase (Phase Difference (PD) equals $\pi$). (online version in color.) }\label{FigApparatusAndSchematic}
\end{figure}

\subsection{Experimental Apparatus}
To experimentally model the hydrodynamic interactions among fish school, we insert a ``schooling" robotic fish into a laminar water tunnel, as shown in Figure. \ref{FigApparatusAndSchematic} (a). The flow velocity of the water tunnel is set as $U=20~ cm/s$, which corresponds to the free-swimming speed of the robotic fish.
The Reynolds number is  $9\times 10^4$, which is similar to the natural situation.
In this article, we study how the Lateral Spaces (LS), Streamwise Spaces (SS) and Phase Differences (PD) affect the power consumption for the tailing fish (see Figure. \ref{FigApparatusAndSchematic} (b)).
Among the three factors, Lateral Spaces (LS) and Streamwise Spaces (SS) are controlled by group of step motors; and phase difference (PD) is controlled by the CPG controller \cite{SM}. 
Referring to the spatial lattices of real fish school \cite{Partridge1980Three-dimensional} and considering the limitation of the width of the flow tunnel, we arrange the lateral space (LS) ranging from 0.27 to 0.33 $BL$, the streamwise space (SS) from 0.35 to 0.45 $BL$ and the phase difference from 0 to 2$\pi$.

As robotic fish is powered by electricity, a measurement and comparison of power consumptions of fish swimming alone and in a formation directly reflect the effects of swimming in groups.
Robotic fish is powered by a  stabilized volt supply (Matrix MPS-3005L-3).
The electrical currents are measured by National Instrument Current Acquisition (NI 9227) with a sample rate of 5000 $Hz$.
The electrical current data acquisition lasts 10 $s$ once after robotic fish swimming stably and repeats 5 times for each spatial lattice and phase difference.

\subsection{Evaluation of power consumption}
Generally, the electrical energy input to the robotic fish will change to mechanical energy and other energy forms such as heat.
For the swimming robot, the useful energy is the mechanical energy which is used to overcomes the reactive force of the fluid.
We evaluate this part of  energy consumption by subtracting the energy consumption when the fish is swimming in air from the energy consumption when fish swim in water \cite{Li2012Hydrodynamic,wen2013Quantitave}.
Thus, the power consumption of fish swimming alone in water is evaluated as $P_{wa} - P_{a}$, where $P_{wa}$ indicates the total power input when the fish swims in water alone, and $P_{a}$ represents the total power consumption while the fish swims in air.
Similarly, we calculate the power consumption for a robotic fish swimming in school by $P_{ws}-P_{a}$, where $P_{ws}$ is the total power consumption when the fish swims in water in school.
To compare the power consumption between fish swimming alone and in groups, we define a power coefficient number $\eta$,
\begin{equation}
  \eta = \frac{(P_{wa}-P_a)-(P_{ws}-p_a)}{P_{wa}-P_a} = \frac{P_{wa}-P_{ws}}{P_{wa}-P_a}
\end{equation}\label{EqEfficiency}
A positive value of $\eta$ indicates a power reduction when fish swimming in school.
On the contrary, a negative value of $\eta$ means fish in school will use more energy.
Notably, all the values are time-averaged as the power consumption varies in one oscillation period.

For each relative position and phase difference between robotic fish, the tests include power acquisitions of robotic fish swimming in air, in water alone and in a school.
Uncertainties during the test quantities are defined as the standard error of the mean $\overline{\sigma}=\sigma/\sqrt{N}$,  for each situation consisting of N =5 measurements.
And  $\sigma$ is the sample standard deviation.

\section{Results}
\subsection{Energy cost depends more on phase difference (PD)}
\begin{figure}[!ht]
  \centering
  \includegraphics[width=0.95\linewidth]{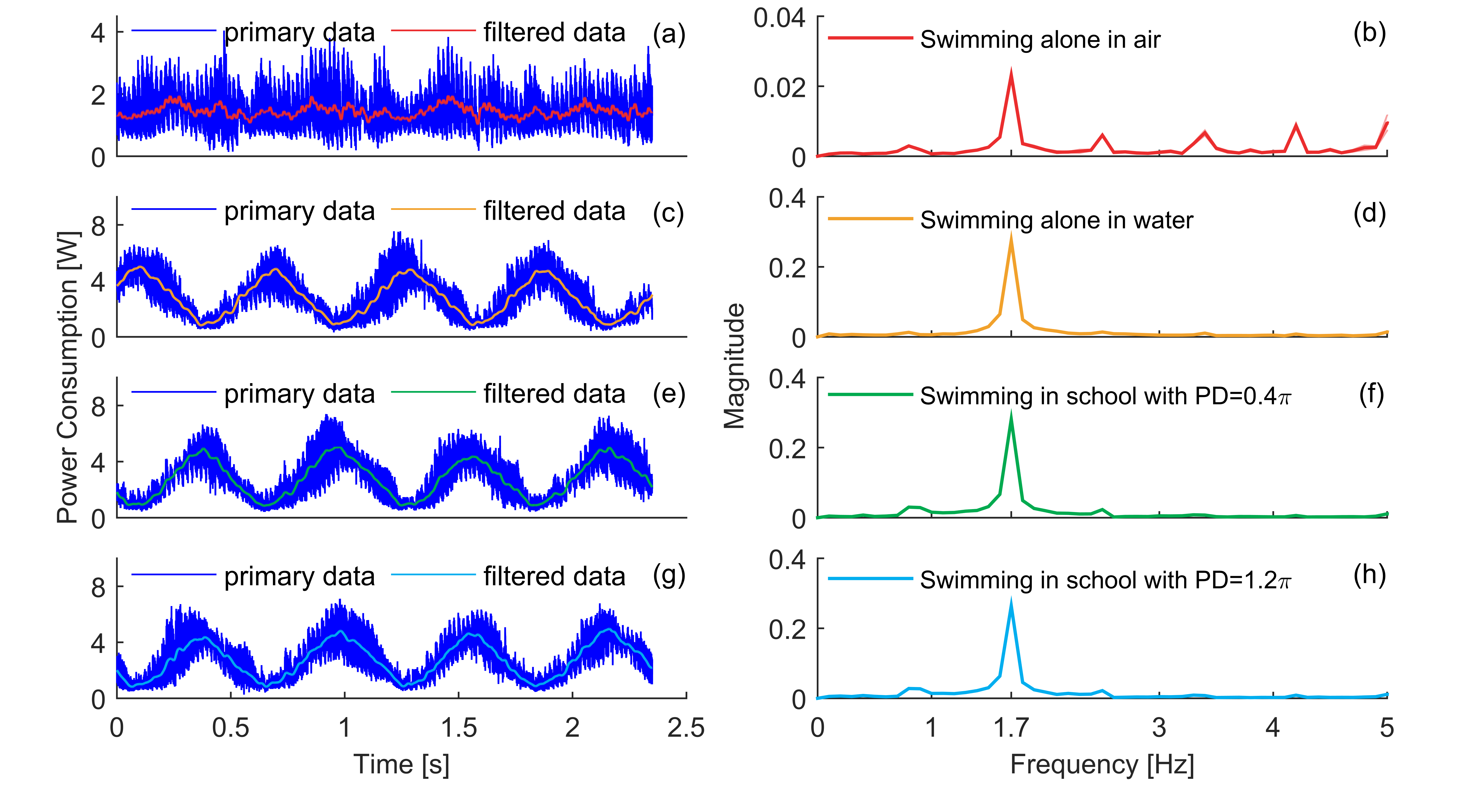}\\
  \caption{Data analysis of power consumptions of the robotic fish. The left column shows the time-domain analysis. And the right column shows the frequency-domain analysis. From the top to the bottom, data in each row are detected in different swimming conditions. (a) and (b), fish swims alone in the air. (c) and (d) fish swims alone in the water. (e) and (f) fish swims in school with $0.4\pi$ PD. (g) and (h) fish swims in school with $1.2 \pi$ PD. }\label{FigDataKalmanAndFFT}
\end{figure}

We first test the performance characteristics of a lone robotic fish both in air and water.
The power consumptions in four different conditions are shown in Figure. \ref{FigDataKalmanAndFFT} (a) (test in air), Figure. \ref{FigDataKalmanAndFFT} (b) (test in water), Figure. \ref{FigDataKalmanAndFFT} (c) (test in school with $0.4\pi$ PD), Figure. \ref{FigDataKalmanAndFFT} (d) (test in school with $1.2\pi$ PD).
Since the power consumption of robot is periodic at twice the frequency of body wave, the power cost is evaluated by the average value.
The average power cost of fish swimming in water $P_{wa} = 2.71 ~W$ is higher than that in air $P_{a} = 1.44~ W$.
Thus when fish swim alone in the water, the power cost to overcome the resistive force of water is $P_{wa}-P_{a} = 1.27~W$.
While fish swim in school with $0.4\pi$ and $1.2\pi$, the power costs to overcome the resistive force of water $P_{ws}-P_a$ are respectively $1.32$ and $1.21$.
Further, the efficiency coefficient for the $0.4\pi$ PD is $-3.94\%$, thus power increase. The efficiency coefficient for the $1.2\pi$ PD is $4.72\%$, thus power reduction.

Note that the noise signal is large in the power values acquired from the experiments.
To verify the effectiveness of the data, we apply the Fourier transform to the data.
As fish undulates the body periodically and the power consumption mainly depends on the reactive force of the fluid,  the power cost value must also be periodic (as the filtered data shown in Figure. \ref{FigDataKalmanAndFFT} (a), (c), (e), (g)).
Since the undulation of fish body is  symmetrical, the frequency of the power $1.7~Hz$ is twice that of the body wave $0.85~Hz$.
Therefore, we prove the effectiveness of the data by this property, as shown in Figure. \ref{FigDataKalmanAndFFT} (b), (d), (f), (h).

%

\begin{figure}[!h]
  \centering
  \includegraphics[width=0.55\linewidth]{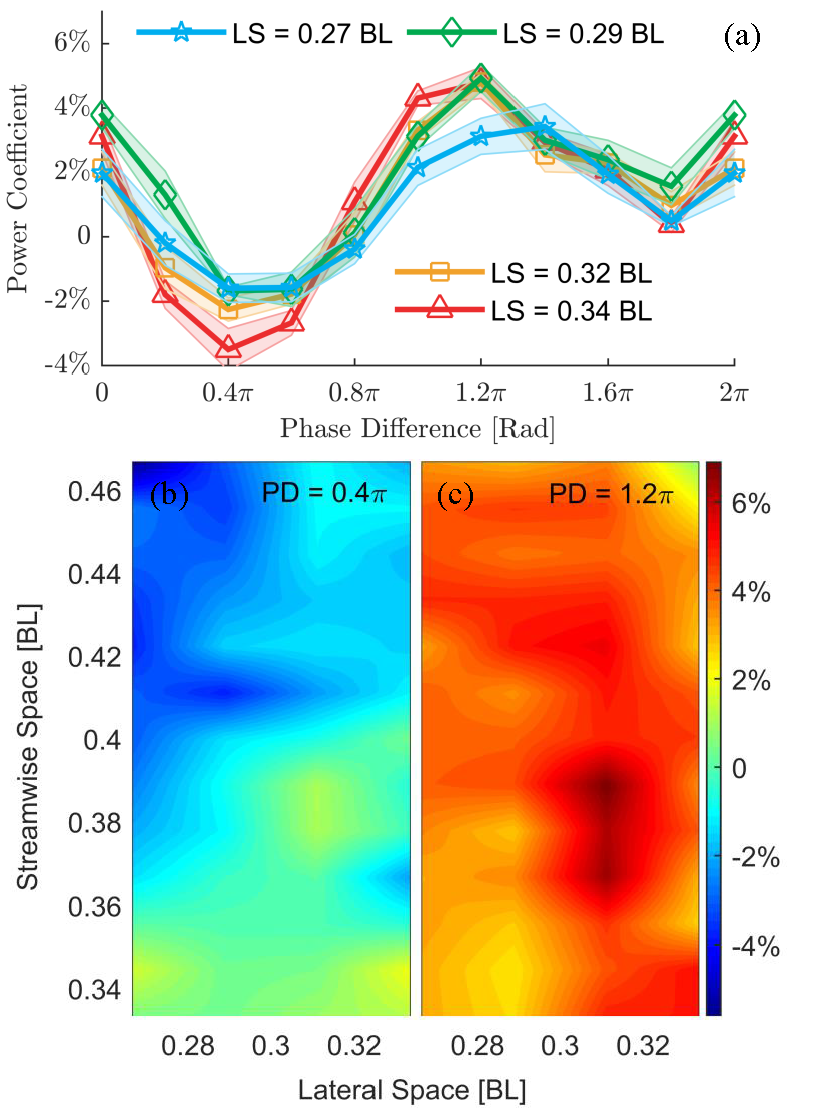}\\
  \caption{Phase difference is critical for fish swimming in groups to extract energy from the aquatic environment.  (a) At the fixed Streamwise Space (SS) 0.4 $BL$ and each Lateral Space (LS), Power coefficient periodically varies across the whole phase. The shadow area shows the stand deviation for 5 trials at each point. (b) The power coefficient versus the spatial spaces at Phase Difference (PD) of 0.4 $\pi$. Most of the coefficients are negative denoting power increase. (c) The power coefficient versus the spatial spaces at Phase Difference (PD) of 1.2 $\pi$. Most of the coefficients are positive denoting power reduction. (online version in color.)}
  \label{FigResultsCollection}
\end{figure}

We then insert a second robotic fish into the flow with a fixed SS of 0.4 $BL$ and series of LS.
The power coefficient is calculated at each spatial lattice as well as phase difference for 5 trials, as shown in Figure. \ref{FigResultsCollection} (a).
The estimated error are illustrated in shadow.
The variation of power coefficient as a function of PD is quite similar to a sine curve and are little related to spatial factors.
Fish school swimming with PD around 0.4 $\pi$ will consume more energy, in schools while around 1.2 $\pi$ schooling will lead energy saving.
Our data show that school fish energy saving is more related to phase differences than spatial lattice.

To prove the hypothesis,  we compare the power variation by fixing the phase difference $0.4\pi$ Figure. \ref{FigResultsCollection} (b), and $1.2\pi$ Figure. \ref{FigResultsCollection} (c).
 In all the areas, fish swimming with phase difference of $0.4\pi$ costs more energy while phase difference of $1.2\pi$ consumes less energy, indicating that the power consumption is more related to the phase difference than the spatial lattices.

\begin{figure}[!ht]
  \centering
  \includegraphics[width=0.7\linewidth]{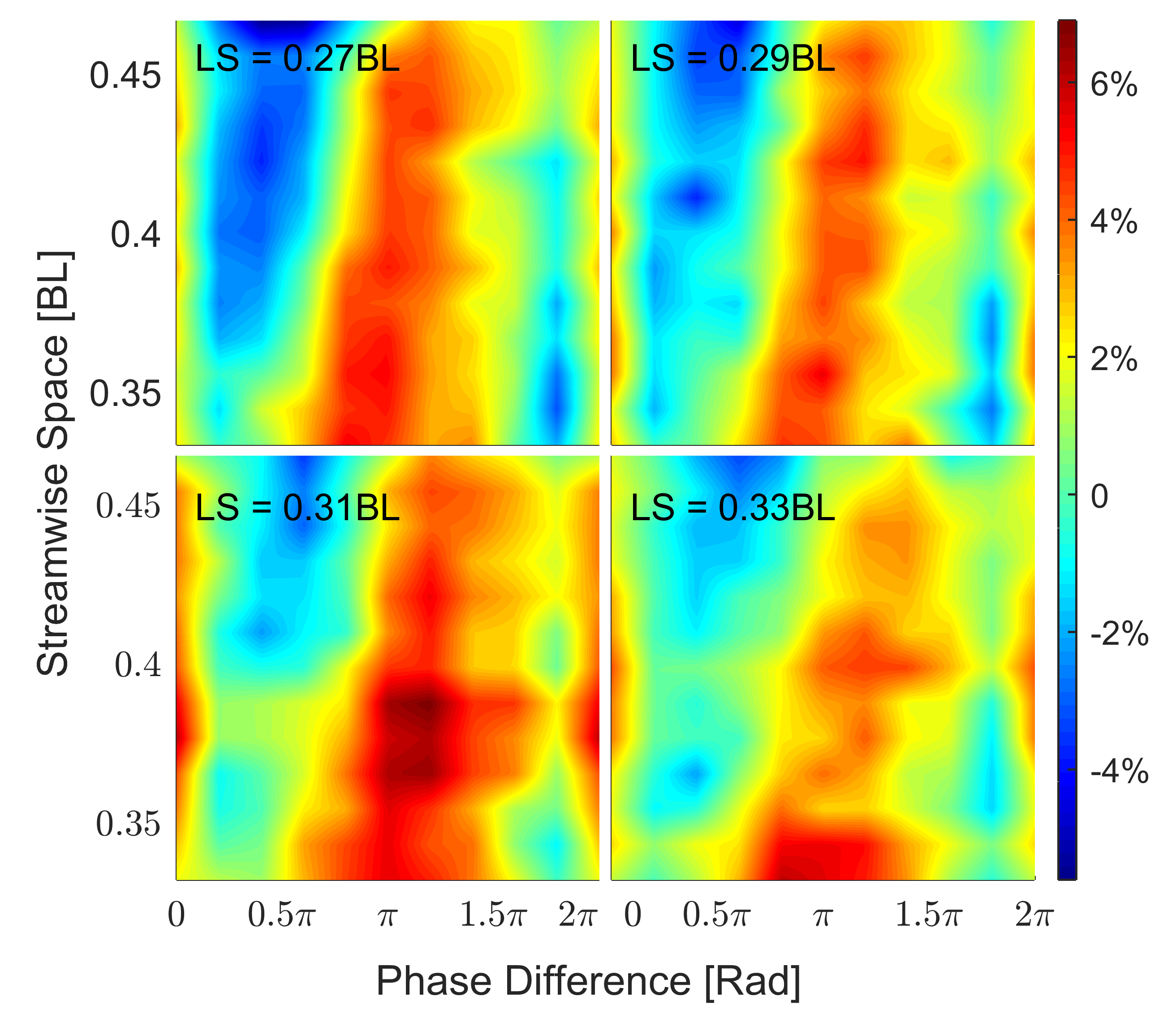}\\
  \caption{Two dimensional map of power coefficient  for the follower swimmer in the (Phase Difference (PD), Streamwise Space (SS)) phase space. Lateral Space (LS) is discretized into (a) $0.27$, (b) $0.29$, (c) $0.31$, (d) $0.33$ (BL). The power reduction and increase are definitely divided by Phase Difference around 0.8 $\pi$. (online version in color.)} \label{FigFollowerPRofPDAndSS}
\end{figure}

To further study how the spatial and temporal factors affect the power consumption of group robotic fish.
We carry out all the situations within the ranges, totaling 2600 trials.
Figure. \ref{FigFollowerPRofPDAndSS} shows the power coefficient as a function of SS and PD with four different LSs.
In contrast to swimming alone, the trailing fish may enjoy energy saving or suffer extra energy cost when swimming in different arrays and Phase Differences (PDs).
For almost all the locations around the leader, the follower will benefit of saving energy when swimming with a phase difference ranging of 0.8$\pi\thicksim$1.4$\pi$.
The larger the SS, the less energy the follower saves.
However, we find that, at larger Streamwise Space (SS), the trailing fish need larger Phase Difference (PD) to get a maximum power reduction.
This indicates that the spatial array also play a role in the power cost of the trailing fish but with less impact than the Phase Difference (PD).
Among all the trails we did here, the maximum power reduction for the follower is 6.9\%, and the power increase is 5.6\%.

\subsection{Hydrodynamic mechanism of energy saving in fish school}

\begin{figure}[!ht]
  \centering
  \includegraphics[width=0.85\linewidth]{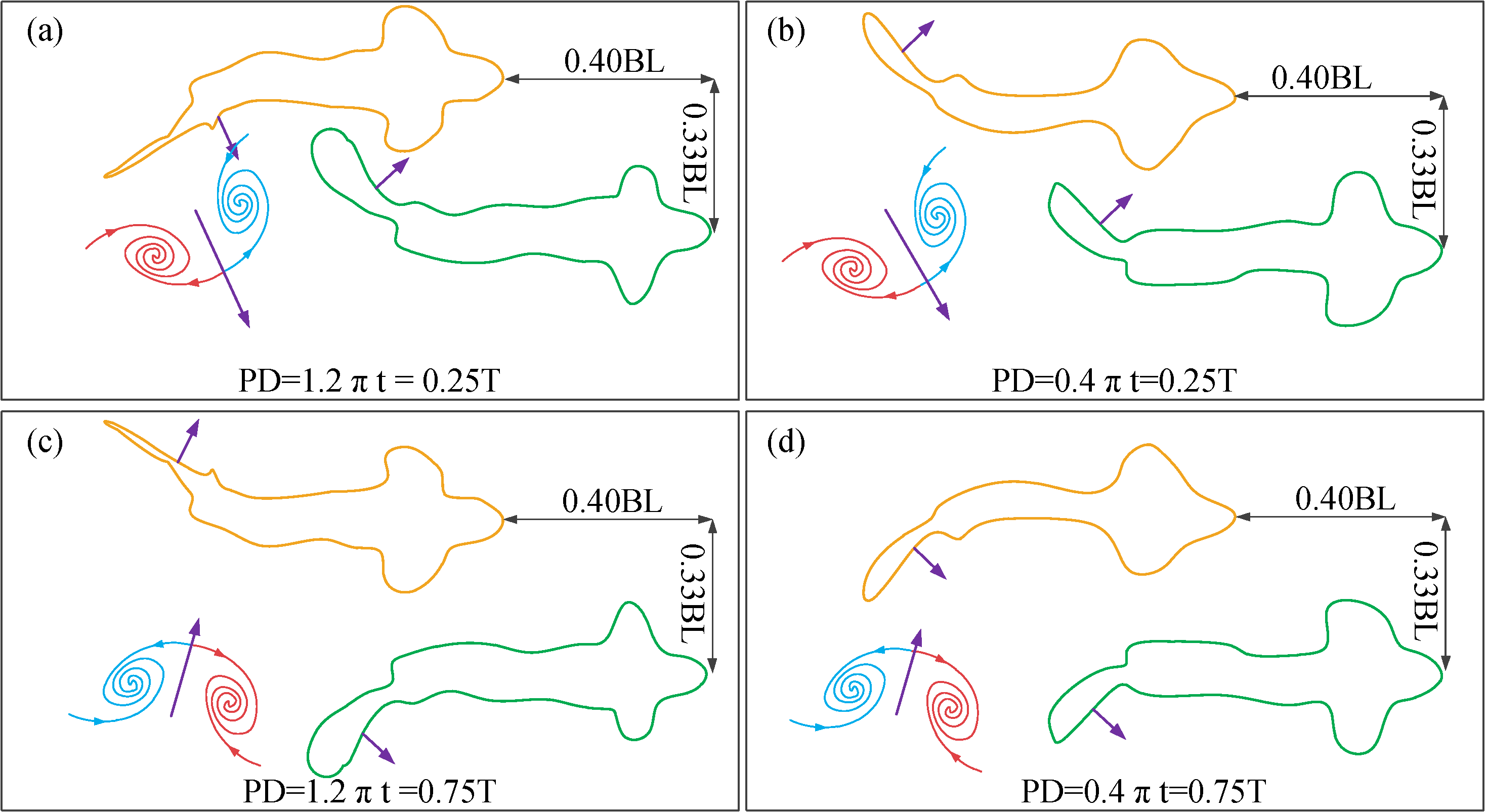}\\
  \caption{ Bottom view of the interactions between the vortices and undulation body of robots. 
  (a), (c) The follower saves energy by flapping in the same direction of the vortex diploe induced by the leader. (b), (d) The follower costs more energy when flapping in the opposite direction of the vortex diploe induced by the leader (See Ref. \cite{SM} for further explanation). (online version in color.)}\label{FigFV}
\end{figure}

To understand why the follower in school saves energy and why the power consumption is strongly related to Phase Difference (PD), we carry out a techchnique of flow visualization to explore the hydrodynamics of fish swimming in school \cite{Smits2000Flow}.
We carry out the experiments by using laser-induced fluorescence (LIF) technique \cite{David1999Unger}.
Since we would like to only show the interactions between the body wave and the shedding vortices, we put some fluorescent species on the fish body and inject the fluorescence dye from the tail.
The fluorescent emission will be captured by the high-speed camera from the bottom of the tunnel at a frequency of 125 Hz (see Figure \ref{FigApparatusAndSchematic}).
As a result, only the parts with fluorescent species will be high lighted; other parts are black (Movies S4, S5).
In order to illustrate the hydrodynamic mechanism of fish school clearly, we show the relationship between the vortices and the body wave in a schematic form in Figure. \ref{FigFV}.
The left column (Figure. \ref{FigFV}(a), (c)) shows why fish in school saves energy in one period.
And the right column (Figure. \ref{FigFV}(b), (d)) denotes the mechanism of why fish in school costs more energy in one period.
The vortex dipole is shed by the leader and have an effect on the follower.
The position of the robotic fish and vortices are all based on the Flow Visualization.

When the direction of the induced vortex dipole is the same as that of the flapping tail, the follower will benefits energy saving (see Figure. \ref{FigFV}(a), (c)).
On the contrary, when the flapping direction is opposite to the direction of the induced vortex dipole, the follower will consume more energy  (see Figure. \ref{FigFV}(b), (d)).
Since PD describes the body wave of the follower leads or lags that of the leader, the relationship between the flapping direction and vortex-dipole direction is also determined by PD.
When the schooling fish swim near (around 0.5~1BL), the energy consumption of the follower is greatly controlled by PD instead of space array.
For all the PDs, PD around $1.2\pi$ leads maximum energy saving for the follower swimming in school, and PD around $0.4\pi$ make the follower consume more energy.

Further, one can definitely find that, with a phase difference of $1.2\pi$, the follower swims in the same direction of the flow field induced by the vortex dipole, thereby saving energy.
While the follower with a Phase Difference of $0.4\pi$ swims in the opposite direction, and thus it costs more energy.
As the flow field induced by the vortex dipole is determined by the leader's flapping tail.
Therefore, the energy cost of the trailing fish essentially depends on the Phase Difference (PD), which proves our hypothesis.

\section{Discussion}

In this article, we applied a high-fidelity robotic fish endowed with a controllable, flexible and active body to investigate the hydrodynamic mechanism of power consumption in a fish school.
Spatial and temporal factors were both considered in the robotic fish school.
The power consumption of the trailing fish is tested and compared when fish are swimming alone and in schools.
Moreover, we consider the spatial arrays that appeared in the real fish school \cite{Partridge1980Three-dimensional}.
And under this spatial condition, we vary the Phase Difference (PD) between the two robotic fish swimming together.
After series of experiments, we find that, under the same spatial condition as in real fish school, the trailing robotic fish saves energy depending more on the PD rather than the array spacing.

Compared with the previous apparatuses, the robotic fish consists of fish-like body, bio-inspired locomotion and similar Reverse Karman Vortex Street (RKVS).
A popular apparatus that was used to uncover the mechanism of energy saving in fish school is the flexible filament in a flowing soap film \cite{Ristroph2008Anomalous,Jia2008Passive}.
As the filament is flexible, the interactions between the body and fluid vortices can be well illustrated \cite{Zhang2000Flexible}.
However, because of the passive body, the vortices shedding by the filaments are Karman Vortices instead of Reverse Karman Vortices.
Using active flapping foils, B. M. Boschitsch  and P. A. Dewey respectively studied the hydrofoils in a in-line configuration \cite{Boschitsch2014Propulsive} and side-by-side \cite{Dewey2014Propulsive} configuration.
Although the rigid foils are active and controllable, They are quite different from the flexible tails, as real fish tail has a  increasing elasticity towards the end.
Moreover, a flapping foil can not represent the hydrodynamics of fish swimming in water, as the flow first encounters the rigid head and then interacts with the flexible body.
The robotic fish we applied here combines both the advantages.
And thus, using  robotic fish to mimic the fish school to explore the mechanism of energy saving is feasible and more realistic.

Note that the maximum energy saving for the follower is only 6.9\%, smaller than previous studies \cite{Jia2008Passive,Ristroph2008Anomalous,Zhu2014Flow-mediated,Boschitsch2014Propulsive}.
One explanation of this is that the real school is more than two fish.
In this article, we only considered the basic schooling of two fish.
As one leader will make the follower save 6.9\%, two leaders will make the follower save about 13.8\%.
Moreover, the frequency of our robotic fish is around 3 times lower than the real fish, therefore the energy of the vortex field shed by our robotic fish is smaller than that shed by the real fish.
This may also lead our data show that the robotic fish only gets a small energy saving.
But in the filed of engineering, an energy saving of 6.9\% is great for the manmade machines.
Our finding in the study may guide the formation and locomotion control for group of bioinspired underwater automatic vehicles.


%
%
The work presented here is basis and needs more work.
The power costs of the leader will be further explored.
Nevertheless, this work first implements artificial fish school, and evaluated the mechanism of fish swimming in groups in the aspect of hydrodynamics.
Further the connections of biology and physics also lie in other aspects such as group behaviour verifications by using bio-inspired high-fidelity robots.

\section*{Acknowledgment}

We thank J. F. Steffensen for supporting the video of real fish.
We also thank W. Wang and C. Wang for helpful discussions.


\end{document}